\documentstyle[prl,aps,multicol,epsf]{revtex}

\begin{document}

\title{On Ground State Tunneling in Mn$_{12}$-Acetate.}
\author{K. M. Mertes and M. P. Sarachik}
\address{Physics Department, City College of the City University of New York,
New York, NY 10031}
\author{Y. Paltiel, H. Shtrikman, and E. Zeldov}
\address{Department of Condensed Mater Physics, 
The Weizmann Institute of Science, Rehovot 76100, Israel}
\author{E. M. Rumberger and D. N. Hendrickson}
\address{Department of Chemistry and
Biochemistry, University of California at San Diego, La Jolla, CA 92093}
\author{G. Christou}
\address{Department of Chemistry, Indiana University, Bloomington, Indiana
47405}
\date{\today}
\maketitle \begin{abstract}
We report Hall sensor measurements of the magnetic relaxation of Mn$_{12}$-acetate 
as a function of magnetic field applied along the easy axis of magnetization for a series of
closely-spaced temperatures between $0.24$~K and $1.9$~K.  We map out a region in the
$H-T$ plane where ground state tunneling is observed, a region where tunneling proceeds via
thermally-assisted tunneling, and a crossover region where both participate in the
relaxation.  We observe the occasional absence of ground-state tunneling under conditions
where one would expect it to be present, and suggest a resolution to the enigma.

\end{abstract}
\vspace {6mm}

\begin{multicols}{2}
Crystals of the high-spin molecular nanomagnet Mn$_{12}$-acetate
([Mn$_{12}$O$_{12}$(CH$_3$COO)$_{16}$(H$_2$O)$_4$]$\cdot$ 2CH$_3$COOH$\cdot4$H$_2$O),
exhibit dramatic quantum mechanical phenomena on a macroscopic scale, and have been the focus
of intense interest and investigation in recent years.  Mn$_{12}$ molecules have recently
been proposed as qubits for quantum computers
\cite{loss,tejada}.  Such qubits would have to be operated at millikelvin temperatures where
the ground state tunneling creates a quantum superposition of spin-up and spin-down
states.  The study of ground state tunneling in Mn$_{12}$ is, therefore, important for
applications.  

The magnetic core of the molecules
consists of twelve Mn atoms strongly coupled by superexchange through oxygen bridges with a
ground-state spin $S=10$ \cite{sessoli2}.  These identical weakly-interacting magnetic
clusters are regularly arranged on a tetragonal body-centered lattice.  As illustrated by
the double well potential shown in the inset to Fig. \ref{fig_steps}, strong uniaxial
anisotropy yields a set of energy levels corresponding to different
projections $m = \pm 10, \pm 9,.....,0$ of the total spin along the easy c-axis of the
crystal.  Measurements\cite{friedman,thomas} below the blocking temperature of 
$3$ K have revealed a series of steep steps in the curves of $M$ versus $H$ at 
roughly equal intervals of magnetic field, as shown in Fig. \ref{fig_steps}, due
to enhanced relaxation  of the magnetization whenever levels on opposite sides of the
anisotropy barrier coincide in energy.  As demonstrated by the data of Fig.
\ref{fig_steps}, different ``steps''  dominate at different temperatures,
indicating that thermal processes play an important role.  The steps in the magnetization
curves have been attributed\cite{novak} to thermally-assisted quantum tunneling of the spin
magnetization.

\vbox{
\vspace{0.4in}
\hbox{
\hspace{-0.1in} 
\epsfxsize 3.3in \epsfbox{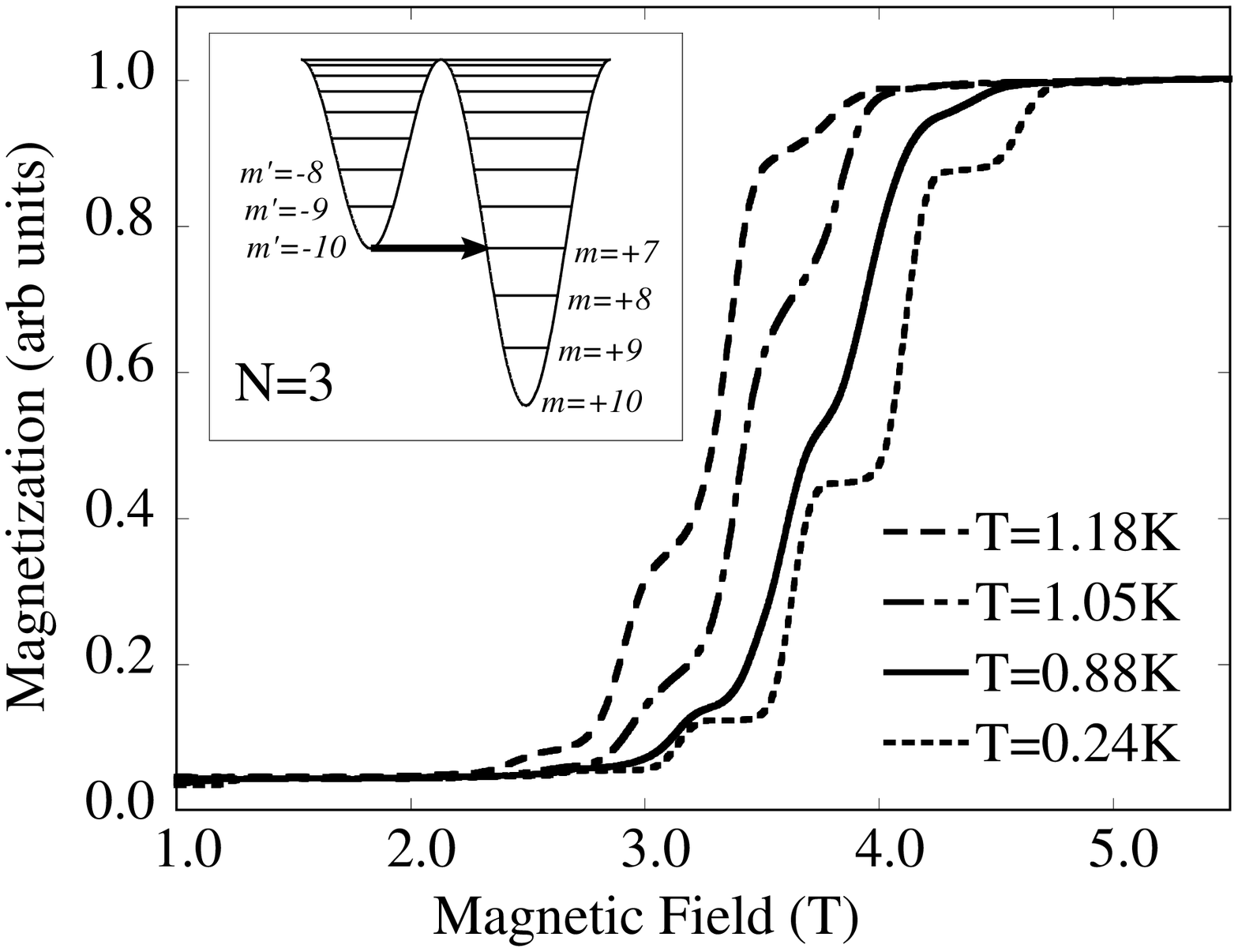} 
}
\vspace{0.2in}
}
\refstepcounter{figure}
\parbox[b]{3.3in}{\baselineskip=12pt FIG.~\thefigure.
Magnetization versus longitudinal magnetic field for a Mn$_{12}$ sample 
starting from a demagnetized state, $M=0$; data are shown at four different 
temperatures, as labeled.  
Note the steep segments, or steps, corresponding to faster magnetic relaxation 
at specific values of magnetic field.
\vspace{0.10in}
}
\label{fig_steps}

\vspace{0in}

Neutron scattering experiments \cite{zhong,hennion,mirebeau} as well as EPR \cite{barra}
measurements indicate that Mn$_{12}$-acetate can be modeled by the effective spin
Hamiltonian:
$$
{\cal H} = -D S_z^2 -g_z \mu_B H_z S_z - A S_z^4 + .......
$$
where $D=0.548(3)$ K is the anisotropy constant, the second term is the Zeeman
energy,  and the third term represents the next higher-order term in longitudinal anisotropy
with $A=1.173(4) \times 10^{-3}$ K; additional small terms that do not commute with the
Hamiltonian and which drive the tunneling (such as transverse internal magnetic  fields,
transverse anisotropy) are not explicitly shown.  Within this model, tunneling should occur
from level
$m'$ in the metastable well to level $m$ in the stable potential well for magnetic fields:
$$
H_z=N\frac{D}{g_z  \mu_B}\left[1 + \frac{A}{D}\left(m^2 + m'^2\right)\right],
$$
where $N = |m + m'|$ is the step number and the level matching condition is $m = -m'-N$. 
The second term inside the bracket is smaller than the first so that steps $N$ occur at
approximately equally spaced intervals of magnetic field, $D/(g_z\mu_B) \approx 0.42$
Tesla.  Careful measurements have revealed structure
within each step due to the presence of the fourth-order term,
$AS_z^4$; the levels do not cross simultaneously, an effect that is more pronounced
for states that are deeper in the well.  This allows identification of the energy levels that
are responsible for the tunneling observed at different temperatures, magnetic fields and
sweep rates.

The process by which the magnetic moment relaxes toward equilibrium depends on temperature:
for temperatures above the blocking temperature, $T_B \approx 3$ K, magnetic relaxation
proceeds by over-the-barrier processes; at intermediate temperatures below $3$ K, 
thermal activation to an excited state within the potential well is followed by tunneling
across the barrier; for temperatures below approximately $1$ K, tunneling in Mn$_{12}$ has
been shown to proceed from the ground state only.  Thermal activation becomes exponentially
more difficult as one proceeds up the ladder to higher energy levels; on the other hand, the
barrier is lower and more penetrable, so that tunneling becomes exponentially
easier.  Which level (or group of adjacent levels) dominates the tunneling is determined by
competition between the two effects.   As the temperature is reduced and thermal activation
becomes more difficult, the  states that are active in the tunneling move gradually to lower
energies deeper in the potential well.

In this paper we report detailed measurements of the magnetization of a
single crystal of Mn$_{12}$ in a swept magnetic field for a set of closely-spaced
temperatures.  Our measurements provide a mapping in the $H-T$ plane of a region where only
thermally-assisted tunneling occurs, a region where only ground state tunneling is observed,
separated by a crossover region where both participate in the relaxation.  We find that
thermally assisted tunneling gives rise to a single broad feature that cannot be
deconvoluted into a set of lines (either Gaussian or Lorentzian) associated with the
different excited spin states in the potential well.  Tunneling from the ground state gives
rise to a separate Gaussian line.  Interestingly, the maxima associated with
thermally-assisted tunneling and ground state tunneling remain separate, while the
contributions associated with the various excited states are unresolved, even though the
field spacing between them is comparable.  Confirming earlier reports
\cite{kent,bokacheva,mertes}, there is an abrupt transfer of ``spectral weight'' to ground
state tunneling as the temperature is reduced.  We show further that under some
circumstances relaxation that should proceed from the ground state appears to be missing
under conditions where one would expect it to be present.  We describe this enigma in
detail, and offer a possible resolution to the puzzle.

\vbox{
\vspace{0.3in}
\hbox{
\hspace{-0.18in} 
\epsfxsize 3.5in \epsfbox{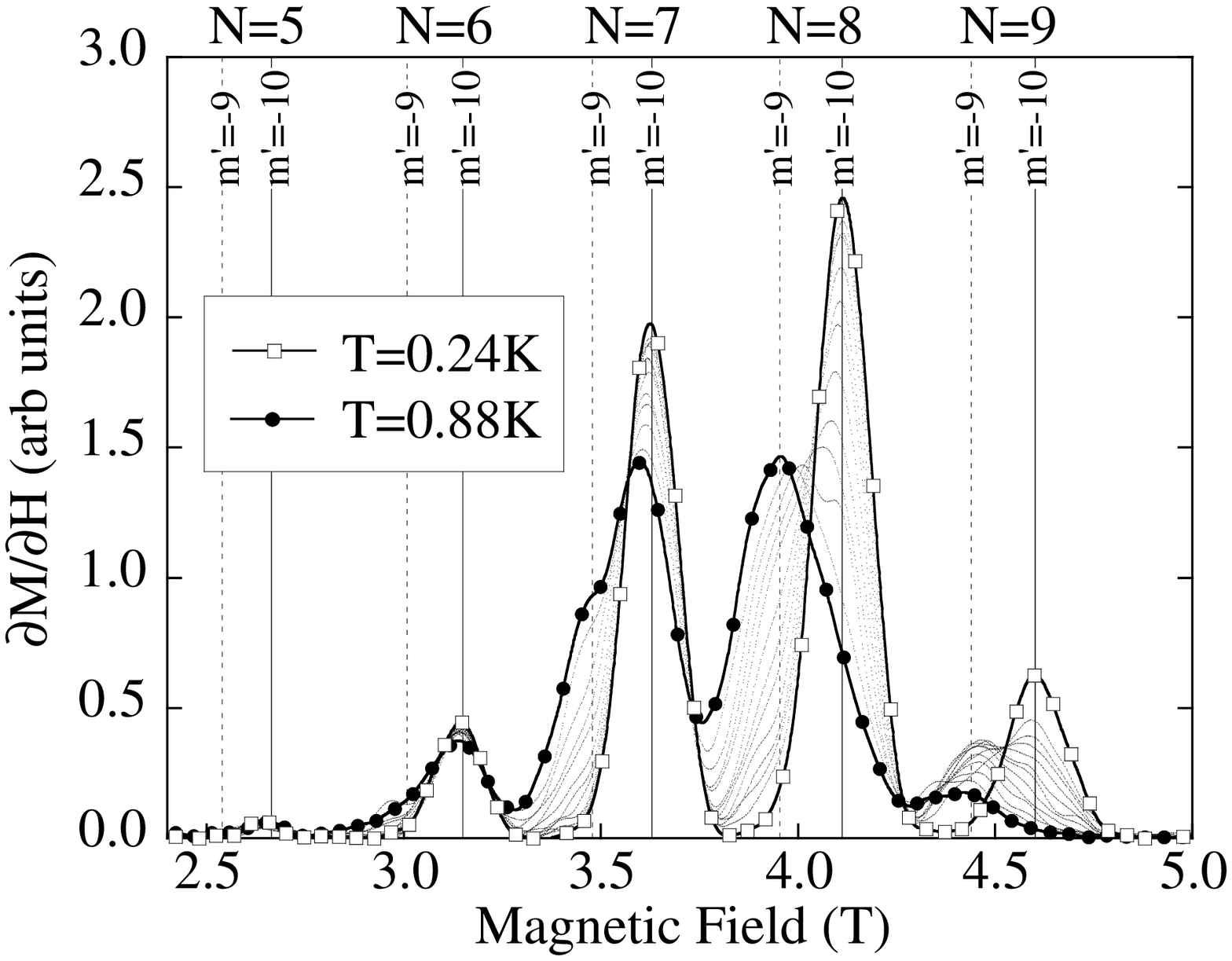} 
}
\vspace{0.2in}
}
\refstepcounter{figure}
\parbox[b]{3.3in}{\baselineskip=12pt FIG.~\thefigure.
For a set of closely spaced temperatures, the first derivative of
the magnetization with respect to magnetic field is plotted as a function of 
magnetic field.  The amplitude is a measure of the rate of magnetic relaxation.  
Selected data points are shown for $0.88$ K and $0.24$ K only.  The remaining curves
(unlabeled) correspond to intermediate temperatures $0.88>T>0.24$ K.  Note the substructure
within each of the four maxima corresponding to steps 
$N= |m'+m|=5, 6, 7, 8,$ and $9$.  
\vspace{0.10in}
}
\label{fig_trans}

The magnetization of small single crystals of Mn$_{12}$ was determined from measurements 
of the local magnetic induction at the sample surface using $10 \times 10$ $\mu$m$^2$ 
Hall sensors composed of a two-dimensional electron gas (2DEG) in a GaAs/AlGaAs 
heterostructure \cite{zeldov}.  The 2DEG was aligned parallel to the external magnetic field,
and  the Hall bar was used to detect the perpendicular component (only) of the magnetic 
field arising from the sample magnetization.  The external magnetic field was swept at a
constant rate of $1.88 \times 10^{-3}$ T$/$s.

For different temperatures 
between $0.24$ K and $0.88$ K, Fig. \ref{fig_trans} shows the first derivative, 
$\partial M/\partial H$, of the magnetization $M$ with respect to the 
externally applied magnetic field $H$\cite{corrections}.  The maxima occur at 
magnetic fields corresponding to faster magnetic relaxation due to level crossings 
on opposite sides of the anisotropy barrier.  In the temperature range of these 
measurements, maxima are observed for $N=|m+m'| =5$ through $9$.  Considerable 
structure associated with different pairs $m, m'$ is clearly seen within each 
step $N$, with a transfer of ``spectral weight'' to higher values of $m'$ deeper in 
the well as the temperature is reduced.  For sufficiently low temperatures,
the curves do not depend on temperature and the tunneling takes place from the lowest, ground
state of the metastable well.

The tunneling appears in the form of two distinct features, one centered about the magnetic
field corresponding to tunneling from the lowest state $m'=-10$ of the metastable well,
and a second feature associated with thermally-assisted tunneling from all the higher energy
levels, $m'= -9, -8, ....$.  We have been unable to separate this broad line into a
superposition of Gaussians or Lorentzians centered at the magnetic fields corresponding to
excited state level crossings.  We note that the differences between the magnetic fields for
the different level crossings within a given step are quite comparable.  For example, for
step $N=7$, tunneling maxima associated with $m'= -10, -9, -8, -7,.....$ are expected at
magnetic fields $3.629, 3.478, 3.352, 3.251....$ T, so that the differences $\Delta H$ are
$0.151, 0.126, 0.101,....$ T.  If one can resolve ground-state tunneling, one should expect
to resolve the excited state levels as well.  Instead,
a single feature corresponding to thermally assisted tunneling remains distinct from
the ground state: it shifts gradually to the right toward higher field as the temperature is
reduced, becomes a shoulder on the low-field side of the ground state peak, and ultimately
merges with it.

The progression can be examined in detail in Fig. \ref{missing6}, where
the derivative of the magnetization with respect to field is shown as a function of magnetic
field for three different temperatures.  For each resonance, $N=...5, 6, 7, 8....$, the two
vertical lines denote the magnetic fields corresponding to tunneling from the first excited
state,
$m'=-9$ (dotted line), and the lowest state, $m'=-10$ (solid line).  At $1.05$ K,
shown in Fig.
\ref{missing6} (a), the tunneling occurs neither from the ground state nor from the first
excited state.  Instead, the three maxima associated with the $N=6, 7, 8$
resonances are probably due to a superposition of tunneling involving thermal
activation to higher states in the well ($m'=-8,-7,..$).  The tunneling at $0.88$ K, shown
in Fig. \ref{missing6} (b), takes place at magnetic fields corresponding to the ground state
for steps $N=5$ and $N=6$, some magnetic relaxation begins to appear at the first excited
state for $N=7$, and for $N=8$ and $9$ there is tunneling only from
$m'=-9$ while ground state tunneling from $m'=-10$ has disappeared.  At the lowest
temperature of $0.24$ K, all tunneling occurs from the ground state in the field range of
these measurements, as shown in Fig. \ref{missing6} (c).

\vspace{0in}
\vbox{
\vspace{0.2in}
\hbox{
\hspace{-0.1in} 
\epsfxsize 3.2in \epsfbox{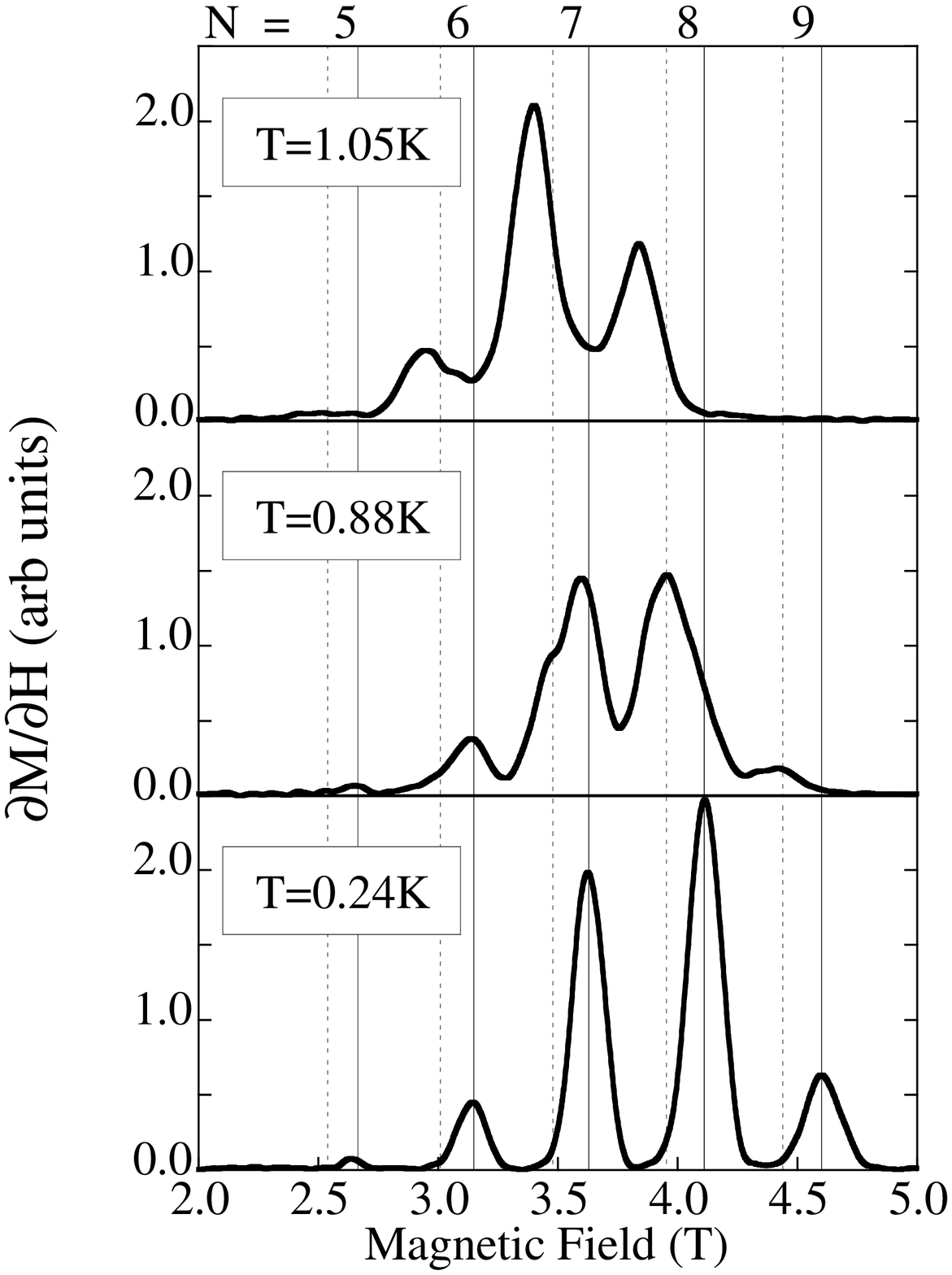} 
}
\vspace{0.1in}
}
\refstepcounter{figure}
\parbox[b]{3.3in}{\baselineskip=12pt FIG.~\thefigure.
The derivative of the magnetization with respect to field versus magnetic field at three
different temperatures, as labeled.  Pairs of vertical lines are drawn for each resonance $N$
corresponding to the magnetic field for tunneling from the first excited state $m'=-9$
(dotted line at the lower field) and from the ground state $m'=-10$ (solid line at the higher
field).
\vspace{0.10in}
}
\label{missing6}

\vspace{0in}

Based on extensive data taken for a closely-spaced set of temperatures, Fig.
\ref{fig_phase} shows a region of the $H-T$ plane where tunneling proceeds from the
ground state only and a region where tunneling occurs from excited states only,
separated by a crossover region where magnetic relaxation involves both ground state and
thermally activated tunneling.  The four points shown in the figure denote the
crossover temperature and field where the two peaks in the
$\partial M / \partial H$ curves have the same amplitude for a given step number,as
illustrated in the inset for step $N=7$.  The line connecting the points in the main
part of Fig. \ref{fig_phase} denotes a boundary (of finite breadth) between thermal
activation and ground state tunneling.  A full mapping of the boundary will require data for
a broader range of sweep rates (slower sweep rates for the high-temperature low-field
region, and faster rates for the high-field low-temperature regime).  Thus, the diagram of
Fig.
\ref{fig_phase} shows that as the magnetic field is increased at a fixed rate at constant
temperature, no ground state tunneling is observed at any magnetic field for temperatures
above $\approx 1.05$ K; at lower temperatures, ground state tunneling is observed for the
lower-numbered resonances, with a shift to thermally-assisted tunneling occuring at higher
magnetic fields.

\vbox{
\vspace{0.2in}
\hbox{
\hspace{-0.25in} 
\epsfxsize 3.6in \epsfbox{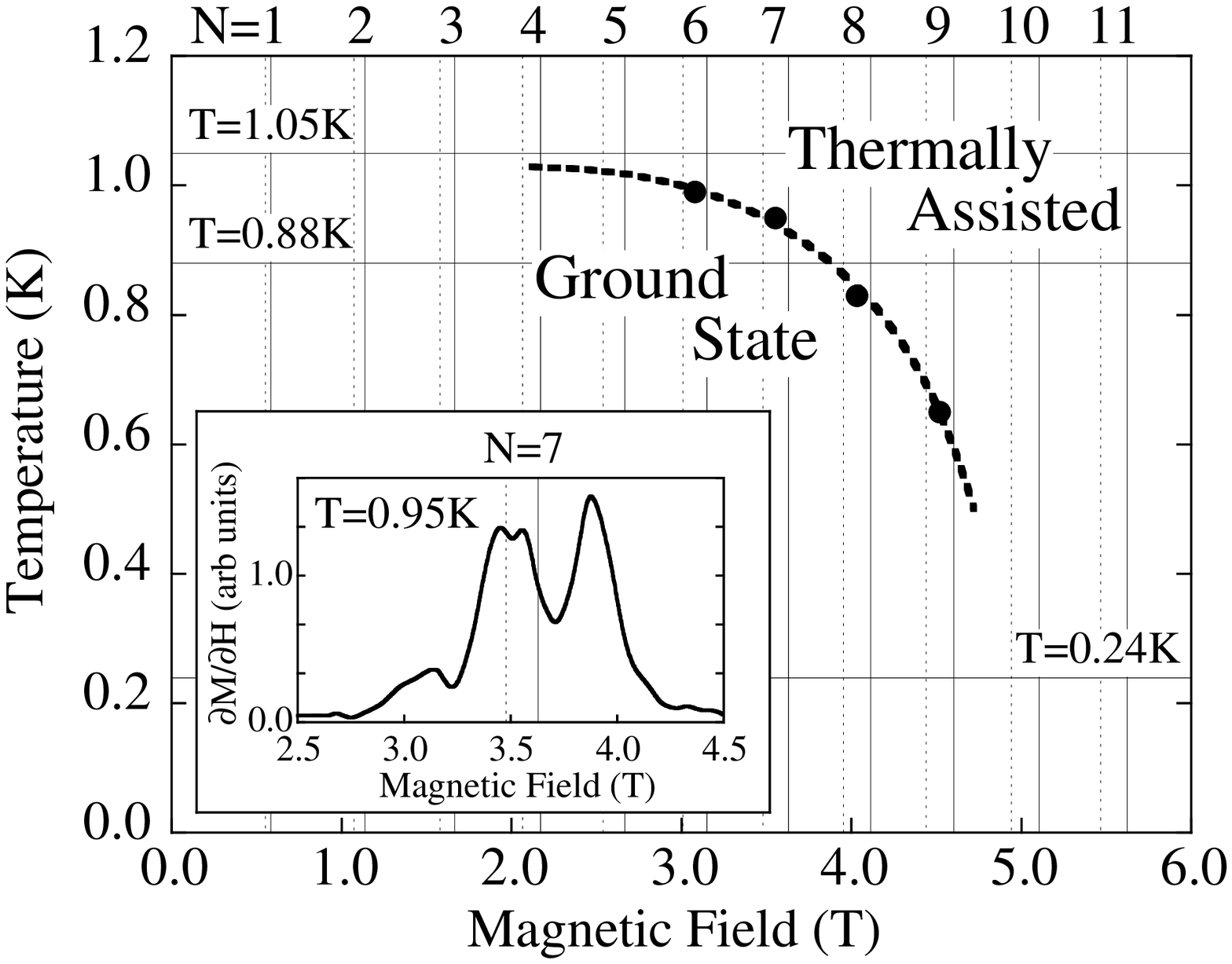} 
}
\vspace{0.1in}
}
\refstepcounter{figure}
\parbox[b]{3.3in}{\baselineskip=12pt FIG.~\thefigure.
Boundary in the $H-T$ plane separating a region where tunneling proceeds from the
ground state only and a region where tunneling occurs from excited states only.  For each
value of $N$, the dotted line denotes the magnetic field for tunneling from the first
excited state, $m'=-9$, and the solid line denotes the field for tunneling from the ground
state, $m'=-10$.  The four points denote the crossover temperature and field where the two
peaks in the $\partial M /\partial H$ curves have the same amplitude for a given step
number; this is illustrated for $N=7$ in the inset.  The line connecting the four points
in the main part of the figure denotes a boundary (of finite breadth) between thermal
activation and ground state tunneling.  Horizontal lines are drawn corresponding to the three
temperatures shown in Figs. \ref{missing6} (a), (b) and (c).
\vspace{0.10in}
}
\label{fig_phase}

\vspace{0in}

We now arrive at the enigma referred to in the introduction.  Examination of resonance $N=7$
at the three temperatures illustrated in Fig. \ref{missing6} shows that tunneling proceeds
entirely from the excited levels at $T = 1.05$ K with no discernible contribution from the
ground state, while the tunneling at the two lower temperatures is entirely due to ground
state tunneling for $N=7$.  The enigma is that ground state tunneling appears to be absent
at the higher temperature.  Similar behavior is found at every other step.  One should bear
in mind that although the population of the excited states is exponentially sensitive to
temperature, $n = n_0 e^{-E/kT}$, the population of the ground state, $n_{gr} = n_0[1 -
e^{-E/kT}] \approx n_0$, at any low temperature, $T<T_B$.  The spin population of the lowest
level in the metastable well is therefore essentially the same at $1.05$ K, $0.88$ K and
$0.24$ K, and if tunneling occurs from the ground state at the lower temperatures, it should
also be observable at $1.05$ K.  One could understand the absence of ground state tunneling
at step
$N=7$ at $1.05$ K if relaxation at lower fields had effectively depleted the
out-of-equilibrium spin magnetization, so that the system has relaxed to near-equilibrium. 
However, as the magnetic field sweeps beyond the field corresponding to ground-state
tunneling at $N=7$ at $1.05$ K, a sizable maximum develops at the next resonance $N=8$,
indicating that an appreciable fraction of the spin magnetization is still out of
equilibrium and is available to relax instead at the next set of level crossings at $N=8$.  

Chudnovsky and Garanin \cite{chudnovsky,garanin} recently suggested that there is a broad
distribution of tunnel splittings in Mn$_{12}$-acetate crystals.  Within their model, this
arises from a locally varying second-order transverse anisotropy which, although forbidden by
tetragonal symmetry in the perfect crystal, is present in real crystals of Mn$_{12}$-acetate
due to dislocations.  The tunneling rate of the magnetic molecules varies from point to
point, with some relaxing very rapidly and others quite slowly.

Regardless of the physical origin, a broad distribution of tunnel splittings provides a
resolution to the enigma discussed above.  At any particular resonance, ($N,m'$), some
fraction of molecular clusters have tunnel splittings that allow them to relax with a
reasonable probability, while other molecules with much smaller tunnel splittings have
relaxation rates that are sufficiently slow that they cannot tunnel.  In the example
discussed above where ground state tunneling is missing at $N=7$, molecules that belong
to the ``fast''-tunneling portion of the distribution relax; if the temperature is
high, they tunnel by thermal activation to excited spin-states, $N=7, m'=-9, -8,...$,
depleting the magnetization of the ``fast''-tunneling  magnetic clusters so that no
magnetization remains that can relax from the ground state $N=7, m'=-10$.  Meanwhile,
the magnetic centers that have small tunnel splittings remain in the metastable
potential well at step $N=7$, and tunnel instead at the next resonance $N=8$ (or higher)
when the magnetic field is now larger and the potential barrier commensurately lower. 
In this way, a broad distribution of
tunnel splittings and tunneling rates provides a natural explanation for the fact that
ground state tunneling is absent in some  circumstances even though a substantial amount
of out-of-equilibrium magnetization remains in the system.

The experimental observation that ground state tunneling is occasionally absent
under conditions when one would expect it to be present was recently analyzed
quantitatively by Garanin and Chudnovsky \cite{garanin} in terms of a dimensionless
parameter $x(m',N,v,T)$ which depends on the resonance $m'$, the step number $N$, the
magnetic field sweep rate $v$, and the temperature $T$.  It was shown in Ref.
\cite{garanin} that for a broad distribution of tunnel splittings the
$m'$ resonance can be observed in a field-sweep experiment only if the condition 
$x(m',N,v,T) < x((m'+1),N,v,T)$ is satisfied.  Otherwise, the molecules which would
cross the barrier at the $m'$ resonance have been depleted at the
$(m'+1)$ resonance.  For a given $m'$ and $N$ this condition may or may not be
satisfied depending on sweep rate $v$ and the temperature $T$.  A detailed comparison
with theory will be published elsewhere.

The enigma of the ``missing'' ground state tunneling provides support for the presence
of  tunnel splittings that vary from molecule to molecule in Mn$_{12}$-acetate.  Ground
state tunneling is strongly affected by such a wide distribution of tunnel rates within the
crystal.  Applications of molecular nanomagnets as qubits will therefore require the use of
isolated individual magnetic molecules.

We are grateful to Yoko Suzuki for discussions, suggestions, and her help with the
figures and the manuscript.  We thank Dmitry Garanin, Eugene Chudvnosky, and Jonathan
Friedman for their valuable input and suggestions on the manuscript.  Work at City
College was supported by NSF grant DMR-9704309 and at the  University of California, San
Diego by NSF grants CHE-0095031 and DMR-0103290.  Support for G. C. was provided by NSF
grant CHE-0071334.  E. Z. acknowledges the support of the German-Israeli Foundation for
Scientific Research and Development.

\end{multicols}


\begin{references}
\bibitem{corrections} Throughout this paper, we have used $H$ instead of 
the total field $B = H + \alpha (4\pi M)$; the field due to the sample 
magnetization is on the order of $300$ Oe.
\end{references}
\end{document}